\newcommand \bs{\begin{subequations}}
\newcommand \es{\end{subequations}}
\newcommand \bea{\begin{eqnarray}}
\newcommand \eea{\end{eqnarray}}
\newcommand \be{\begin{equation}}
\newcommand \ee{\end{equation}}

\documentclass[
aps,prl, amssymb, showpacs, twocolumn ]{revtex4}

\usepackage{graphicx}
\usepackage{amsmath}
\usepackage{SIunits}

\begin{document}

\title{Size-Dependence of the Wavefunction of Self-Assembled Quantum Dots}

\author{Jeppe Johansen$^1$}
\email{jjo@com.dtu.dk}
\author{S\o ren Stobbe$^1$}
\author{Ivan S. Nikolaev$^{2,3}$}
\author{Toke Lund-Hansen$^1$}
\author{Philip T. Kristensen$^1$}
\author{J\o rn M. Hvam$^1$}
\author{Willem L. Vos$^{2,3}$}
\author{Peter Lodahl$^1$}
\email{pel@com.dtu.dk}

\affiliation{$^1$COM$\cdot$DTU, Department of Communications,
Optics, and Materials, Nano$\cdot$DTU, Technical University of
Denmark,
DTU - Building 345V, DK-2800 Kgs. Lyngby, Denmark\\
$^2$Center for Nanophotonics, FOM Institute for Atomic and Molecular Physics (AMOLF), Amsterdam, The Netherlands \\
$^3$Complex Photonics Systems, MESA+ Institute for Nanotechnology,
University of Twente, The Netherlands}

\pacs{42.50.Ct, 78.67.Hc, 78.47.+p }

\begin{abstract}
  The radiative and non-radiative decay rates of InAs quantum dots are
  measured by controlling the local density of optical states
  near an interface. From time-resolved measurements we extract the oscillator
  strength and the quantum efficiency and their dependence on emission
  energy. From our results and a theoretical model we
  determine the striking dependence of the overlap of the electron and hole
  wavefunctions on the quantum dot size. We conclude that the optical quality is best for large quantum
  dots, which is important in order to optimally tailor
  quantum dot emitters for, e.g., quantum electrodynamics experiments.
\end{abstract}

\maketitle

Semiconductor quantum dots (QDs) have attracted significant
attention recently as nano-scale light sources for
\emph{all-solid-state} quantum electrodynamics experiments
\cite{Reitmaier,Yoshie,Peter,Lodahl1,Kress,Englund}. Major
advancements have culminated in the demonstration of strong
coherent coupling between a single QD and the optical mode of a
cavity \cite{Reitmaier,Yoshie,Peter}. The coupling strength
between the emitter and the cavity is determined by the
\emph{oscillator
  strength}, which is an intrinsic property of the emitter. For atomic
transitions the oscillator strength attains only discrete values
depending on the choice of atom and is determined by the
electrostatic potential. In contrast, the QD oscillator strength
can be ingeniously tailored due to the influence of
size-confinement on the electron-hole wavefunction
\cite{AndreaniPRB1999,Hours05,Finley}. Consequently the oscillator
strength can be continuously tuned by varying the size of the QDs.
Surprisingly the exact size-dependence of the optical properties
of the exciton has remained an open question. Understanding these
excitonic optical properties is much required in order to
optimally engineer QDs for enhanced light-matter interaction.

In this Letter we present measurements of the oscillator strength
of the ground-state exciton in self-assembled QDs. The detailed
dependence on the QD size is mapped out by time-resolved
measurement of spontaneous emission at different emission
energies. We employ the modified local density of optical states
(LDOS), caused by reflections in a substrate-air interface, to
separate radiative and non-radiative decay contributions. This
method was pioneered by Drexhage for dye molecules \cite{Drexhage}
and used also to extract the quantum efficiency of Erbium ions
\cite{Snoeks} and colloidal nanocrystals \cite{Brokmann,Walters}.
Here we use this method to accurately determine the dependence of
the oscillator strength on the quantum dot size. The precise
measurements of the oscillator strength allows us to determine the
size-dependence of the electron and hole wavefunction overlap.

\begin{figure}
  \includegraphics[width=\columnwidth]{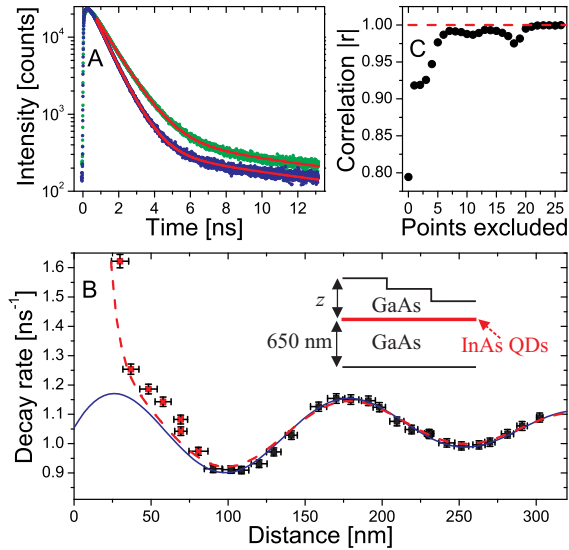}
   \caption{(color online). \textbf{A} Decay of the spontaneous emission recorded at $1.20 \,\mathrm{eV}$ for two different distances to the interface
     of $z=109\,\mathrm{nm}$ (green, upper curve) and
     $z=170\,\mathrm{nm}$ (blue, lower curve).  The bi-exponential fits
     (solid red lines) result in $\Gamma_{\mathrm{f}} =
     0.91\,\mathrm{ns}^{-1},$ $\Gamma_{\mathrm{s}} =
     0.09\,\mathrm{ns}^{-1}$ for $z=109\,\mathrm{nm}$ and
     $\Gamma_{\mathrm{f}} = 1.15\,\mathrm{ns}^{-1},$
     $\Gamma_{\mathrm{s}} = 0.10\,\mathrm{ns}^{-1}$ for
     $z=170\,\mathrm{nm}$.  The goodness-of-fit parameters $\chi ^2_r$
     are respectively $1.17$, and $1.11$, close to the ideal
     value of unity \cite{Lakowicz} verifying the bi-exponential
     model. \textbf{B} Measured decay rates versus distance $z$ to the
     GaAs-air interface (dots).  Calculated LDOS projected onto a
     dipole orientation parallel to the interface (solid blue line).
     Calculated LDOS including dissipation at the surface (dashed red
     line).  The inset is a schematic drawing of the sample.
     \textbf{C} Coefficient of correlation versus the number of data
     points excluded in the modeling of the data in \textbf{B}.}
\label{fig:decay_analysis}
\end{figure}

Time-resolved spontaneous emission is measured from a series of
samples that contain identical ensembles of InAs QDs positioned at
controllable distances to a GaAs-air interface.  The wafer is
grown by molecular beam epitaxy on a GaAs (100) substrate where
$2.0$ monolayers of InAs are deposited at $524 ^{\circ} \,
\mathrm{C}$ followed by a $30 \,\mathrm{s}$ growth interrupt and
deposition of a $300 \,\mathrm{nm}$ thick GaAs cap. The QD density
is $250 \,\micro \mathrm{m}^{-2}.$ A $50 \,\mathrm{nm}$ thick
layer of AlAs is deposited $650 \,\mathrm{nm}$ below the QDs for
an optional epitaxial lift-off. The wafer is processed by standard
UV-lithography and wet chemical etching, whereby samples with
different distances between the QDs and the interface are
fabricated on the same wafer, see insert of
Figure~\ref{fig:decay_analysis}B. The distances $z$ from the
QD-layers to the interface are measured by a combination of
secondary ion mass spectroscopy and surface profiling with typical
precisions of $ \pm 3.0 \,\mathrm{nm}$.

The QDs are excited by optical pumping of the wetting layer states
at $1.45 \,\mathrm{eV}$ using $\sim\! 300 \,\mathrm{fs}$ pulses
from a mode-locked Ti:sapphire laser. The excitation spot has a
diameter of $\sim\!  250 \,\micro\mathrm{m}$ and the excitation
density is kept at $7 \,\mathrm{W/cm}^2$ with a repetition rate of
$82 \: \mathrm{MHz}.$ Under these conditions less than $0.1$
excitons per QD are created, i.e., only light from the QD ground
state is observed. The spontaneous emission is collected by a lens
(NA=0.32), dispersed by a monochromator, and directed onto a
silicon avalanche photo diode for time-correlated single-photon
counting \cite{Lakowicz}. The detection energy is varied between
$1.17 \,\mathrm{eV}$ and $1.27 \,\mathrm{eV}$ to probe different
sub-ensembles of the inhomogenously broadened ground state.  The
spectral resolution of the monochromator is $2.6\,\mathrm{meV},$
which is narrow relative to the bandwidth of the LDOS changes. The
time-resolution of the setup is $48 \,\mathrm{ps}$ given by the
full width half maximum of the total instrument response function.
All measurements are performed at $14 \,\mathrm{K}$.

Figure~\ref{fig:decay_analysis}A shows the spontaneous emission
decay for QDs positioned at two different distances from the
GaAs-air interface and recorded at an emission energy of $1.20
\,\mathrm{eV}$. A clear change in the decay curve is observed with
distance to the interface. The decay of the QD ground state is
very well modeled as a bi-exponential decay, $I(t)=A_{\mathrm{f}}
e^{-\Gamma_{\mathrm{f}}t}+ A_{\mathrm{s}}
e^{-\Gamma_{\mathrm{s}}t} + C$, over the complete time range of
the measurement. The background level $C$ is determined by the
measured dark count rate and after-pulsing probability of the
detector. The fast decay takes place on a time scale of about $1
\,\mathrm{ns}$ corresponding to the decay of bright excitons in
InAs QDs. The slow decay time is approximately $10 \,\mathrm{ns}$
and does not systematically depend on the distance and is most
likely due to recombination of dark excitons \cite{Favero}. In the
remainder of this Letter we will focus only on the fast decay
rate.

The decay rates measured at $1.20 \,\mathrm{eV}$ are presented in
Figure~\ref{fig:decay_analysis}B as a function of distance from
the QDs to the interface. A damped oscillation of the total decay
rate with distance is observed. The data are compared to the LDOS
calculated for GaAs (assuming $n=3.5$) and projected onto a dipole
orientation parallel to the interface (solid blue line). Only the
parallel component is relevant since refraction in the interface
reduces the solid angle for light collection, and additionally the
QD orientation is predominantly parallel to the interface
\cite{Cortez}. The measured decay rate $\Gamma(\omega,z)$ is the
sum of a non-radiative $\Gamma_{\mathrm{nrad}}(\omega)$ and a
radiative $\Gamma_{\mathrm{rad}}(\omega,z)$ decay rate. The latter
is proportional to the projected LDOS $\rho(\omega,z)$ and depends
explicitly on the distance $z$ to the interface and the optical
frequency $\omega$. It is calculated as the sum over all available
electromagnetic modes projected onto the orientation of the
dipole, and is obtained using a Green's function approach
\cite{CPS}. We define the radiative decay rate for QDs in a
homogeneous medium $\Gamma_{\mathrm{rad}}^{\mathrm{hom}}(\omega)$
and express the measured total decay rate as
\begin{equation}\label{eq:linearized_problem}
\Gamma_{}(\omega,z) = \Gamma_{\mathrm{nrad}}(\omega) +
\Gamma_{\mathrm{rad}}^{\mathrm{hom}}(\omega)
\frac{\rho(\omega,z)}{\rho_{\mathrm{hom}}(\omega)},
\end{equation}
where $\rho_{\mathrm{hom}}(\omega)$ is the LDOS of a homogeneous
medium of GaAs.

Excellent agreement between experiment and theory is observed in
Figure~\ref{fig:decay_analysis}B for distances of $z \ge 75\,
\mathrm{nm}$. This explicitly confirms the validity of the
theoretical model used to extract properties of the emitter as
opposed to a previous work \cite{Brokmann}. For QDs closer than
$75\, \mathrm{nm}$ to the GaAs-air interface the measured decay
rates are systematically larger than the calculated rates. We
exclude that this effect is due to tunneling out of the QDs, which
has been observed only within
 $15\, \mathrm{nm}$ from a surface \cite{Wang}. An
increased non-radiative loss may be due to scattering or
absorption at the surface of the etched samples. This dissipation
in the surface is modeled as a thin absorbing surface layer, which
creates an optical surface state. The dashed line in
Fig.~\ref{fig:decay_analysis}B is obtained by including a
$5\,\mathrm{nm}$ thick layer with refractive index of $3.5 + 1.0 i
$, which leads to increased rates near the interface in agreement
with the experimental data.

For QDs sufficiently far away from the interface the influence of
any surface effects is negligible, and our data can be used to
reliably extract QD properties. We determine the data points that
are not influenced by the dissipation in the surface as follows:
Eq. (\ref{eq:linearized_problem}) reveals a linear relation
between the measured rate and the calculated normalized LDOS. We
therefore perform a linear regression analysis and obtain the
linear correlation parameter $|r|$ as the data close to the
interface are excluded point by point, cf.
Figure~\ref{fig:decay_analysis}C.  After excluding the seven
closest data points the correlation parameter converges to unity,
hence Eq.  (\ref{eq:linearized_problem}) is valid.  By comparing
experiment and theory we determine the radiative and non-radiative
decay rates at $1.20 \,\mathrm{eV}$ to be
$\Gamma_{\mathrm{rad}}^{\mathrm{hom}} =
0.95\pm0.03\,\mathrm{ns}^{-1}$ and $\Gamma_{\mathrm{nrad}} =
0.11\pm 0.03\, \mathrm{ns}^{-1}$, which is the most accurate
result to date for QDs.

\begin{figure}
  \includegraphics[width=\columnwidth]{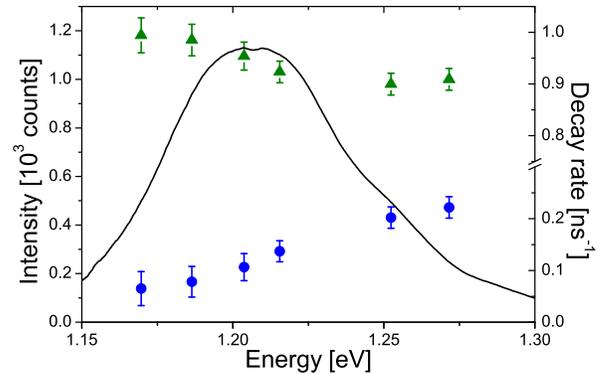}
   \caption{(color online). Left axis: Photoluminescence from the
     inhomogenously broadened ground state measured at $z=281\,\mathrm{nm}$ (solid line). Right axis: Radiative (green triangles) and
     non-radiative (blue circles) decay rates versus emission energy.}
\label{fig:sixEnergies}
\end{figure}

The measurements have been performed for six different energies within
the inhomogenously broadened emission spectrum of the QDs. The
inhomogeneous broadening reflects the different sizes of QDs such that
small QDs correspond to high emission energies and \textit{vice
  versa}. The radiative and non-radiative decay rates extracted from
the measurements are plotted in Fig.~\ref{fig:sixEnergies}. The
increased non-radiative recombination rate at higher energies could
indicate that carriers can be trapped at the QD surface since the
relative importance of the surface is large for small QDs. While such
a size dependence would be general for all QDs, the absolute values of
the non-radiative rates could depend on sample growth. Surprisingly
the radiative rate is found to decrease with increasing energy. This
behavior is due to the decrease of the overlap between the electron
and hole wavefunctions as the size of the QD is reduced, as discussed
below. Our method allows to extract the size-dependence of the QD
emission without any implicit assumption of vanishing non-radiative
recombination, as opposed to previous works \cite{Hours05,vanDriel}.
In fact such an assumption would in our case lead to the incorrect
conclusion of an increased wavefunction overlap with reduced QD size.

Fermi's Golden Rule relates the radiative decay rate and the
oscillator strength
\begin{equation}\label{eq:Fermis_homogenous}
f_{\mathrm{osc}}(\omega)= \frac{6 m_\mathrm{e} \epsilon_0 \pi
c_0^3 }{q^2 n \omega^2} \times
\Gamma_{\mathrm{rad}}^{\mathrm{hom}}(\omega),
\end{equation}
where $n$ is the refractive index of GaAs, $\omega$ is the
frequency of the optical transition, $m_\mathrm{e}$ is the
electron mass, $\epsilon_0$ is the vacuum permittivity, $q$ is the
elementary charge, and $c_0$ is the speed of light in vacuum. For
the QDs emitting at $\hbar \omega = 1.20 \,\mathrm{eV},$ the
measured value of $\Gamma_{\mathrm{rad}}^{\mathrm{hom}}$ results
in an oscillator strength of $f_{\mathrm{osc}}=13.0\pm0.4$. For
comparison various estimates of the oscillator strength based on
absorption measurements have been reported in the literature and
are generally in the range of $f_{\mathrm{osc}}$=5-10
\cite{Birkedal,Warburton}. However, the technique implemented here
provides unprecedented precision since it only relies on accurate
measurements of the distance of the QDs to the interface and is
independent of, e.g., the QD density. Additionally, the quantum
efficiency, i.e., the ratio of the radiative decay rate to the
total decay rate, is extracted. We find $QE =90 \pm 4\%$ at the
emission energy of $1.20 \,\mathrm{eV}$, which confirms that a
high quantum efficiency is feasible with an ensemble of emitters
and not only with selected single QDs \cite{Brokmann}. The
intrinsically high QD quantum efficiency can be further increased
by tuning the size as discussed below.

\begin{figure}
  \includegraphics[width=\columnwidth]{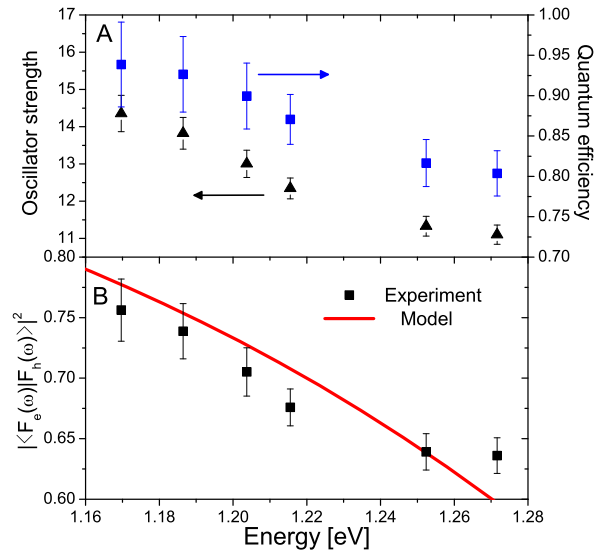}
  \caption{(color online). \textbf{A} Oscillator strength (triangles) and quantum efficiency (squares) versus energy. \textbf{B} Measured (squares) and calculated (red curve) energy dependence of the overlap of the envelope wavefunctions of electrons and holes. }
\label{fig:dipolemoments}
\end{figure}

The energy dependence of the oscillator strength and the quantum
efficiency are presented in Figure~\ref{fig:dipolemoments}A. Both
quantities are seen to decrease with increasing energy. The
quantum efficiency decreases from around $95\%$ to $80\%$ and the
oscillator strength from $14.5$ to $11$ over the inhomogenously
broadened emission spectrum. This result shows that large QDs with
a high exciton confinement potential have much better optical
properties than smaller QDs. Our results thus shed new light on
the optimum design of solid-state QED experiments, and strong
coupling was indeed observed for large QDs
\cite{Reitmaier,Yoshie,Peter}.

Our measurements provide new insight into the size dependence of
the QD's wavefunctions. Thus, the spatial overlap between the
electron and hole wavefunctions $\left|
\Psi_{\mathrm{e,h}}(\omega) \right>$ can be obtained from the
oscillator strength.
%
%
Within  the effective mass approximation, valid in the strong
confinement limit when Coulomb effects are negligible, the
electron (or hole) wavefunction can be factorized in a conduction
(or valence) band Bloch wavefunction for InAs $\left|
u_{\mathrm{c/v}}\right>$, and an electron (or hole) envelope
function $\left| F_{\mathrm{e/h}}(\omega)\right>$. The overlap of
the electron and hole wavefunctions is related to the oscillator
strength via \cite{DelerueBook,Singh,vanDriel}
\be |\langle F_\mathrm{e}(\omega)|F_\mathrm{h}(\omega) \rangle|^2
= \frac{m_\mathrm{e} \hbar \omega }{6  |\langle
u_\mathrm{v}|\mathbf{\hat{e}}\cdot\mathbf{p}|u_\mathrm{c}
\rangle|^2} \times f_\mathrm{osc}(\omega), \label{eq:overlap}\ee
where $\mathbf{\hat{e}}$ is the polarization unit vector of the
electromagnetic field and $\mathbf{p}$ is the electron momentum.
Evaluation of the matrix element is performed as an average over
all possible orientations of the polarization vector. As strain
lifts the degeneracy of the light-hole and heavy-hole bands, only
transitions from the conduction band to the heavy-hole band are
included. The result can be expressed as $ |\langle
u_\mathrm{v}|\mathbf{\hat{e}}\cdot\mathbf{p}|u_\mathrm{c}
\rangle|^2= m_\mathrm{e}E_\mathrm{p}/6$, where $E_\mathrm{p}=22.2
\,\mathrm{eV}$ is the Kane energy for bulk InAs \cite{Singh}. From
Eq. (\ref{eq:overlap}) and the measured radiative decay rates, the
wavefunction overlap is obtained, see
Fig.~\ref{fig:dipolemoments}B. The wavefunction overlap decreases
from about $0.75$ at $1.17 \,\mathrm{eV}$ to $0.63$ at $1.27
\,\mathrm{eV}.$ The reduction in the oscillator strength stems
from the increased mismatch between the electron and hole
wavefunctions with decreasing QD size. The reduction of the
wavefunction overlap is due to the more sensitive size-dependence
of the electron wavefunction compared to the hole wavefunction due
to their difference in effective mass. As the size of the QD is
decreased, the electron will penetrate deeper into the barrier
than the hole thus reducing the overlap \cite{Narvaez2005}. The
detailed understanding of how the size affects the overlap between
the electron and hole wavefunctions is crucial in order to
optimally tailor QDs for efficient coupling to light.

The observed energy dependence of the oscillator strength is compared
to a simple effective-mass QD model. We use finite-element-method
calculations to obtain the energy levels and the corresponding
wavefunctions of the electron and holes. The curve in
Fig.~\ref{fig:dipolemoments}B displays the wavefunction overlap
calculated for a lens-shaped QD with a radius of $7\,\mathrm{nm}$ and
a height varying between $1.8\,\mathrm{nm}$ and $3.0\,\mathrm{nm}$.
The following realistic parameters are used: a wetting layer thickness
of $0.3\,\mathrm{nm}$, $60\%$ of the band-edge discontinuity is in the
conduction band, and the GaAs content in the QDs is taken to be
$25\%$. Good agreement with the experimental data is observed, and the
theory clearly confirms a pronounced reduction of the electron-hole
wavefunction overlap as the size of the QD is decreased. The general
validity has been tested by calculating for different sizes, shapes,
and amount of GaAs content in the QD, and they all show a decrease of
overlap with increasing energy. The same behavior is also obtained
from more involved QD models also in the presence of strain
\cite{Andreev2005}.  Interestingly, the radiative rate was observed to
increase with energy for colloidal nanocrystals in agreement with
theory \cite{vanDriel}, which is opposite to the results reported here
for InAs QDs. This illustrates a striking difference in the optical
properties of colloidal nanocrystals compared to self-assembled QDs,
which is due to their different sizes and confinement potentials.

In summary, we have measured the radiative and non-radiative decay
rates of InAs QDs by employing the modified LDOS near a dielectric
interface. The oscillator strength and quantum efficiency of the
QDs and their dependence on the emission energy were accurately
determined. The radiative decay rate decreases with increasing
energy leading to a reduction of the oscillator strength. In
contrast the non-radiative recombination rate increases with
increasing energy corresponding to a reduction of the quantum
efficiency. Consequently QDs emitting on the low-energy side of
the inhomogenously broadened ground state transition are most
suitable as nanophotonic light sources due to their optimized
optical properties. The experimental findings are explained by a
model of the QD taking the size-dependence of the wavefunctions
into account. Our results demonstrate how QD wavefunctions can be
tailored to achieve improved coupling to light, which is needed in
order to take full advantage of the potential of quantum
electrodynamics devices based on QDs.

We thank Martin Olesen, Torben Kristensen and Stig Salomonsen for
supplying the program code for the QD-model and Mads Lykke
Andersen and Ad Lagendijk for stimulating discussions. We
gratefully acknowledge the Danish Research Agency (division FNU)
for financial support. This work was part of the EU project
"Qphoton". ISN and WLV are supported by FOM and NWO-Vici.

\end{document}